\documentclass[twocolumn,showpacs,aps,pre,amsmath,amssymb]{revtex4}
\usepackage{graphicx}
\usepackage{dcolumn}
\usepackage{bm}

\begin{document}

\title{Refraction of Electromagnetic Energy for Wave Packets Incident on a
Negative Index Medium is Always Negative}

\author{W. T. Lu, J. B. Sokoloff, and S. Sridhar}
\address{Physics Department and Electronic Materials Research Institute,
 Northeastern University, Boston, MA 02115.}

\date{\today }

\begin{abstract}
We analyze refraction of electromagnetic wave packets on passing from an
isotropic positive to an isotropic negative refractive index medium. We
definitively show that in all cases the energy is always refracted
negatively. For localized wave packets, the group refraction is also always
negative.
\end{abstract}


\maketitle

The existence of a medium with a negative ($n<0$) index
of refraction (NIM), raised several years ago \cite{Veselago}, has been
demonstrated experimentally recently \cite{Shelby}. 
One of the most striking properties of NIM's is that of
negative refraction for plane waves across the interface between positive
index materials (PIM) and NIM. Negative refraction means that when radiation
passes through an interface between a PIM and an NIM, the refracted beam is
on the same side of the normal as the incident beam (see Fig. 1), in
contrast to the usual positive refraction in which they are both the opposit
sides of the normal. 

In studies of negative refraction, 
it is essential to represent incident waves as localized wave packets, rather
than plane waves, since all physical sources of electromagnetic waves
produce radiation fields of finite spatial and temporal extent because the
sources are always of finite spatial extent and because they only radiate
for a finite time.  Hence treatments of this problem which study waves that 
extend over infinite distance in all or some directions cannot be trusted to
reliably predict the direction in which a wave will be refracted, and in
fact treatments based on such extended waves \cite{Valanju} have led to a
direction of refraction opposite that which one finds for spatially
localized wave packets, resulting in a great deal of controversy and confusion. 
Although several treatments using  waves 
of infinite extent in some direction (e.g., a plane wave front \cite{pacheco}%
) have  obtained negative refraction, since such a
model is unphysical, for the reasons given above, we cannot have confidence
in conclusions obtained from it. 

In this article, we treat  refraction of a localized wave packet at a PIM-NIM
interface both analytically and by simulations, demonstrating that it 
refracts negatively. We also present both analytic and numerical studies 
of wave packets constructed from a small number of plane waves, on the basis 
of which we are able to give a plausible explanation for why the two plane wave 
model studied by Valanju, et. al.\cite{Valanju}, gives the wrong answer. We 
find that in all cases, including the model of Valanju, et. al., the energy 
and momentum of the wave refract negatively. Since electromagnetic waves are 
detected only when they either give up energy to or exert a force on a detector, 
the relevant direction of propagation to consider is that of the region of 
space in which the energy and momentum of the wave are nonzero.

Without sources, Maxwell's equations are $\nabla \cdot \mathbf{D}=0$, $%
\nabla \times \mathbf{H}=\partial _{t}\mathbf{D}$, $\nabla \times \mathbf{E}%
+\partial _{t}\mathbf{B}=0$, $\nabla \cdot \mathbf{B}=0$. For plane waves of 
wave vector \textbf{k} and and frequency $\omega $, 
only three equations are independent. Using the usual relationships between 
{\bf D}(t) and {\bf E}(t) and between {\bf B}(t) and {\bf H}(t)\cite{Landau} 
one obtains for such plane waves 
$\mathbf{k}\times \mathbf{H}=-\omega \varepsilon (\omega)\mathbf{E}$, 
$\mathbf{k}\times \mathbf{E}=\omega \mu (\omega) \mathbf{H}$. Combining 
these equations gives us a functional relationship between $\omega$ and 
{\bf k}. Wave propagation is only permitted for (%
$\varepsilon ,\mu ,n>0$) or ($\varepsilon ,\mu ,n<0$) \cite{Veselago}. In
the later case, $(\mathbf{E},\mathbf{H},\mathbf{k})$ will form a left-handed
triplet while for an ordinary material, $(\mathbf{E},\mathbf{H},\mathbf{k})$
form a right-handed triplet.

A wave packet localized in a compact region of space, as occurs in all
experimental situations, can be constructed from a continuous distribution
of wavevectors. Consider such a wave packet incident from outside the NIM, $%
\mathbf{E}=\hat{\mathbf{y}}E_{0}\int d^{2}kf(\mathbf{k}-\mathbf{k}_{0})e^{i(%
\mathbf{k}\cdot \mathbf{r}-\omega (\mathbf{k})t)}$ with $\omega (\mathbf{k}%
)=ck$. Here we only consider $S$-polarized waves. The $P$-polarized waves
can be treated similarly, however. Throughout the paper, we choose the 
$z$-axis from
PIM to NIM normal to the interface and the $x$-axis along the interface. If $f(%
\mathbf{k}-\mathbf{k}_{0})$ drops off rapidly as $\mathbf{k}$ moves away
from $\mathbf{k}_{0}$, $\omega (\mathbf{k})$ can be expanded in a Taylor
series to first order in $\mathbf{k}-\mathbf{k}_{0}$ to a good
approximation. This gives, $\mathbf{E}=\hat{\mathbf{y}}E_{0}e^{i(\mathbf{k}%
_{0}\cdot \mathbf{r}-\omega (\mathbf{k}_{0})t}g(\mathbf{r}-ct\mathbf{k}%
_{0}/k_{0})$, with $g(\mathbf{R})=\int d^{2}kf(\mathbf{k}-\mathbf{k}%
_{0})e^{i(\mathbf{k}-\mathbf{k}_{0})\cdot \mathbf{R}}$.

Inside the NIM, $\mathbf{k}$ and $\mathbf{k}_{0}$ in the argument of the
exponent get replaced by $\mathbf{k}_{r}$ and $\mathbf{k}_{r0}$ which are
related to $\mathbf{k}$ and $\mathbf{k}_{0}$ by Snell's law 
\begin{equation}
k_{rx}=k_{x},\quad k_{rz}=-\sqrt{(n_{r}\omega /c)^{2}-k_{x}^{2}}.
\end{equation}%
Here $n_{r}$ is the refractive index for the NIM and is a function of $%
\omega $. Then the wave packet once it enters the NIM is given by 
\begin{equation}
\mathbf{E}_{r}=\hat{\mathbf{y}}E_{0}e^{i(\mathbf{k}_{r0}\cdot \mathbf{r}%
-\omega (\mathbf{k}_{r0})t)}g_{r}(\mathbf{r}-\mathbf{v}_{gr}t),
\end{equation}%
where $g_{r}(\mathbf{R})=\int d^{2}kf(\mathbf{k}-\mathbf{k}_{0})t_{\mathbf{k}%
}e^{i\mathbf{R}\cdot (\mathbf{k_{r}}-\mathbf{k_{r0}})}$ and where $t_{%
\mathbf{k}}$ is the transmission amplitude for an incident plane wave of
wave vector $\mathbf{k}$. It is the standard expression for this quantity
for the two polarizations of the incident plane wave\cite{jackson} except
that the index of refraction in each medium is replaced by the index of
refraction divided by the permeability of the medium. The latter
modification of the expressions follows quite simply from the arguments
given in that reference, if one does not replace the ratio of the
permeabilities by 1, as was done in Ref. 6. Here $\mathbf{k}_{r0}$ denotes $%
\mathbf{k}_{r}$ evaluated at $\mathbf{k}=\mathbf{k}_{0}$ and $\mathbf{v}%
_{gr}=\nabla _{\mathbf{k}_{r}}\omega (\mathbf{k}_{r})$ evaluated at $\mathbf{%
k}_{r}=\mathbf{k}_{r0}$. Let us expand $\mathbf{k_{r}}-\mathbf{k_{r0}}$ in
the exponential function in the expression for $g_{r}(\mathbf{R})$ in a
Taylor series in $\mathbf{k}-\mathbf{k_{0}}$ to first order, 
$\mathbf{k_{r}}-\mathbf{k_{r0}}\approx 
({\bf k}-{\bf k_0})\cdot
\nabla_{\bf k}({\bf k_r}-{\bf k_{r0}})|_{{\bf k}={\bf k_0}}.$
Substituting this in the expression for $g_r({\bf r})$, we obtain $g_{r}(%
\mathbf{R})=\int d^{2}kf(\mathbf{k}-\mathbf{k}_{0})t_{\mathbf{k}}e^{i\mathbf{%
R}\cdot \lbrack (\mathbf{k}-\mathbf{k_{0}})\cdot \nabla _{\mathbf{k}}(%
\mathbf{k_{r}}-\mathbf{k_{r0}})]}$. If the width of the distribution of wave
vectors $f(\mathbf{k}-\mathbf{k_{0}})$is small compared to the range of k
over which $t_{\mathbf{k}}$various significantly, we can to a good
approximation simply evaluate this quantity at $\mathbf{k}=\mathbf{k}_{0}$ 
and put it outside the integral over $\mathbf{k}$. Then, the transmission
coefficient of the wave packet is simply given by $|t_{\mathbf{k_{0}}}|^{2}$.

If we carry out the expansion of $\omega (\mathbf{k}_{r0})$to second order
in $\mathbf{k}-\mathbf{k}_{0}$, we are able to show that the wave packet
spreads out, but if the length and width of the packet are much larger than
the wavelength corresponding to the wave vector $\mathbf{k}_{0}$ at the the
peak in $f(\mathbf{k}-\mathbf{k}_{0})$, we find that the amount that the
packet spreads out in a given time interval is much smaller than the
distance traveled by the packet in that time. Then clearly under such
reasonable conditions, the wave packet will remain sufficiently well-defined
to be able to observe the refraction of the packet. The expansion of the
frequency in a Taylor series is valid for a sufficiently narrow distribution 
$f(\mathbf{k}-\mathbf{k}_{0})$.

In order to get an explicit expression for $g(\mathbf{R})$, let the wave
packet have a Gaussian form $f(\mathbf{k})=(\Delta x\Delta z/\pi)\exp[
-k_x^2(\Delta x)^2-k_z^2(\Delta z)^2]$. Expanding $\mathbf{k}_r$in a Taylor
series around $\mathbf{k}_{r0}$, we get 
\begin{equation}
g_r(\mathbf{R})=\exp[-{C_x^2/4(\Delta x)^2}-{C_z^2/4(\Delta z)^2}]
\end{equation}
with $C_x=R_x+({cn_r/\upsilon_r}-1)(k_{x0}/k_{rz0})R_z$, $%
C_z=(cn_r/\upsilon_r)(k_{z0}/k_{rz0})R_z$, and $\upsilon_r=c(dn_r
\omega/d\omega)^{-1}$. From the above expressions, one can see that the
Gaussian wave packet moves with $\mathbf{v}_{gr}$. Due to the dispersion,
the wave packet is deformed in the NIM.

An NIM is dispersive and causality demands that $\mathrm{d}(\varepsilon
\omega)/ \mathrm{d}\omega>1$ and $\mathrm{d}(\mu \omega)/\mathrm{d}\omega>1$
for nearly transparent media \cite{Landau,SmithKroll}. For an isotropic NIM, 
since $n_r$ is a function of $\omega $ only, $\mathbf{v}%
_{gr}=c(dn_r\omega/d\omega)^{-1}(c\mathbf{k} _r/n_r\omega)=-\upsilon_r{\hat {%
\mathbf{k}}}_r$ with $\hat{\mathbf{k}}_r$ the unit vector in the 
direction of $\mathbf{k}_r$.
Since $\upsilon_r$ is always positive for transparent media as required by
causality, the group velocity will be refracted opposite the direction of 
wavevector $\mathbf{k}_r$.

The magnetic field obtained from the electrical field through $%
\mathbf{H}=(1/\omega \mu )\mathbf{k}\times \mathbf{E}$ is%
\begin{equation}
\mathbf{H}_{r}=-{\frac{E_{0}}{c}}\int d^{2}kf(k-k_{0}){\frac{n_{r}(k)}{\mu
_{r}(k)}}(\hat{\mathbf{k}}_{r}\times \hat{\mathbf{y}})e^{i\mathbf{k}%
_{r}\cdot \mathbf{r}-i\omega (k)t}.
\end{equation}%
from which we find the Poynting vector to be
\begin{eqnarray}
\mathbf{S}_{r} &=&-{\frac{E_{0}^{2}}{c}}\int d^{2}k\int d^{2}k^{\prime }f(%
\mathbf{k}-\mathbf{k}_{0})f(\mathbf{k^{\prime }}-\mathbf{k}_{0})\hat{\mathbf{%
k}}_{r}  \nonumber \\
&&\times {\frac{n_{r}(k)}{\mu _{r}(k)}}\cos \left[ \mathbf{k}_{r}\mathbf{-}%
\omega (k)t\right] \cos \left[ \mathbf{k^{\prime }}_{r}\cdot \mathbf{r}%
-\omega (k^{\prime })t\right] ,
\end{eqnarray}%
where we have used the fact that $\mathbf{k}_{r}\cdot \hat{y}=0.$ While
there is no question that the Poynting vector at a point in a medium gives
the local direction of energy flow, it does not give us the direction of
energy flow by a wave packet or a group of plane waves as a whole since the
direction of the Poynting vector varies with space. The integral of the
Poynting vector over all space, $\mathbf{P}_{r}=\int \mathbf{S}_{r}d\mathbf{r%
}$, however, gives the total momentum carried by a wave packet. This
quantity divided by the volume over which the wave packet is nonzero is the
average of the Poynting vector over the whole wave packet. Either way, this
integral clearly represents the direction of motion of the wave packet in
the medium. From the above expression for $\mathbf{S}_{r}$, one has 
\begin{equation}
\mathbf{P}_{r}=-(E_{0}^{2}/2c)\int d^{2}kf(\mathbf{k}-\mathbf{k}_{0})^{2}%
\hat{\mathbf{k}}{\frac{n_{r}(|k|)}{\mu _{r}(|k|)}}.
\end{equation}%
Let us consider a coordinate system whose $z$-axis is along $\mathbf{k}_{0}$%
. The function $f(\mathbf{k}-\mathbf{k}_{0})^{2}$ will then be a function of 
$k_{x}$ and $k_{z}$ symmetrically peaked around $k_{x}=0$ and $k_{z}=k_{0}$.
Then writing Eq. (6) as 
\[
\mathbf{P}=-{\frac{E_{0}^{2}}{2c}}\int d^{2}kf(\mathbf{k}-\mathbf{k}%
_{0})^{2}{\ \frac{k_{x}\hat{\mathbf{x}}+k_{z}\hat{\mathbf{z}}}{|k|}}{%
\frac{n_{r}(|k|)}{\mu _{r}(|k|)}}
\]%
we can see that since k is an even function of $k_{x}$, the integrand is an
odd function of $k_{x}$ and hence the $x$-component vanishes. Therefore, $%
\mathbf{P}$, which as argued above represents the propagation direction of
the wave packet, is opposite in direction to $\mathbf{k}_{0}$, i.e., in the
direction of the group velocity. Hence, the energy refracts negatively.

\begin{figure}[htbp]
\center{
\includegraphics [angle=0,width=3.6in]{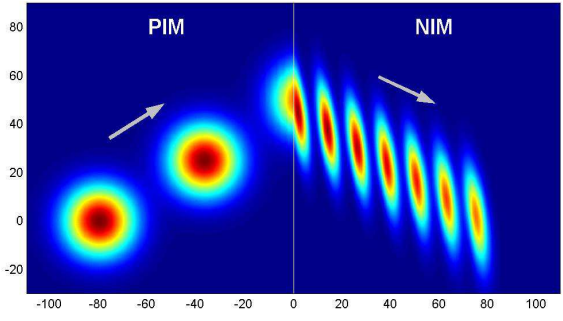}}
\vskip -0.8cm
\caption{Time-lapse snap shots of the electric field intensity of a
propagating Gaussian wave packet refracting negatively at a PIM-NIM
interface. The center wave number is $k_0=\protect\sqrt{5}$ with incident
angle $\protect\pi/6$. The spatial extent of the incident wave packet is $
\Delta x=\Delta z=10$. The time step is 50 with speed of light $c=1$. The
dispersion Eq. (\protect\ref{disp}) were used for NIM and $n=\protect\mu =1$
for PIM.}
\label{Fig1}
\end{figure}

The negative refraction of the wave packet is illustrated by numerical
simulation in Fig.\ref{Fig1}. We use the following dispersion relation 
\begin{equation}
n_{r}(\omega )=-(1/\omega )\sqrt{(\omega ^{2}-\omega _{b}^{2})(\omega
^{2}-\omega _{p}^{2})/(\omega ^{2}-\omega _{0}^{2})}.  \label{disp}
\end{equation}%
for the NIM with $\omega _{0}<\omega <\omega _{b}$. The permeability is $\mu
_{r}=(\omega ^{2}-\omega _{b}^{2})/(\omega ^{2}-\omega _{0}^{2})$. The
numbers we used in the calculation are, $\omega _{0}=1$, $\omega _{b}=3$, 
$\omega _{p}=\sqrt{10}$, $c=1$. Fig.\ref{Fig1} shows stroboscopic snapshots
of the electric field intensity of a propagating wave packet incident on a
PIM-NIM interface. The negative refraction of the wave packet is clearly
evident.

\begin{figure}[htbp]
\center{
\includegraphics [angle=0,width=3.6in]{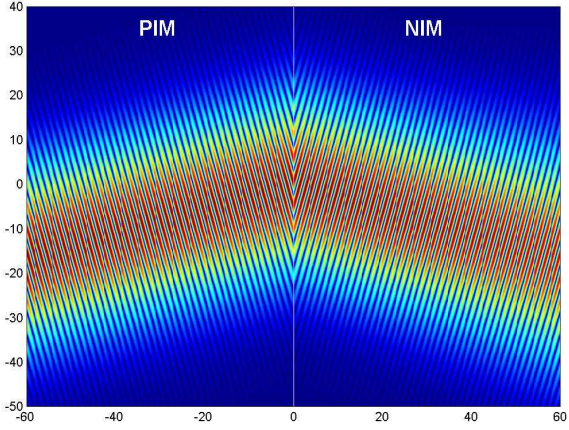}}
\vskip -.8cm
\caption{Electric field $\Re E$ of a beam with $k_0=\protect\sqrt{5}$ and
the Gaussian weight $f(k_{\bot})=e^{-(10k_{\bot})^2}$. The incident angle of
the beam is $\protect\theta=\protect\pi/12$.}
\label{Fig2}
\end{figure}

A beam can be constructed as follows: 
\begin{equation}
E=E_0\int dk_{\bot}e^{i(\mathbf{k}_0+\mathbf{k}_{\bot})\cdot \mathbf{r}
}f(k_{\bot}).
\end{equation}
Here $\mathbf{k}_{\bot}$ is perpendicular to $\mathbf{k}_0$ and $f(k_{\bot})$
assumes a Gaussian form. Note that this construction is different from that
of Kong et al \cite{Kong02} and Smith et al \cite{Smith02} in that the width
of the incident packet is made finite in directions perpendicular to the
direction of propagation. Because the NIM is highly dispersive, the incident
beam once it enters the NIM will no longer be a beam. It will be a localized
wave packet instead. The electric field $E$ of the beam is shown in Fig.\ref
{Fig2}. Just as for the wave packet, the beam intensity also refracts
negatively.

We next consider the refraction of wave packets made up of a finite number
of plane waves. For the cases of 2 and 3 plane waves analytical expressions
are obtained for the Poynting vector, momentum and group velocities. First
consider the case of two plane waves in the $xz$-plane incident from PIM to
NIM where the interface is at $z=0$. Let wave vectors and frequencies be $(%
\mathbf{k}_{1},\omega _{1})$ and $(\mathbf{k}_{2},\omega _{2})$. We set the
polarization in the $y$-direction. Suppose $\Delta \omega =\omega
_{2}-\omega _{1}>0$. The incident wave in PIM is $E=2E_0 e^{i\mathbf{K}\cdot 
\mathbf{r}-i\Omega t}\cos (\Delta \mathbf{k}\cdot \mathbf{r}/2-\Delta \omega
t/2)$ with $\mathbf{K}=(\mathbf{k}_{1}+\mathbf{k}_{2})/2$, $\Omega =(\omega
_{1}+\omega _{2})/2$, and $\Delta \mathbf{k}=\mathbf{k}_{2}-\mathbf{k}_{1}$ 
and where $E_0$ is the wave amplitude, The electric field of the refracted 
waves is 
\begin{equation}
\mathbf{E}_{r}=2E'_0 e^{i\mathbf{K}_{r}\cdot \mathbf{r}-i\Omega t}\cos
(\Delta \mathbf{k}_{r}\cdot \mathbf{r}/2-\Delta \omega t/2)\hat{\mathbf{y}}
\label{2-plane}
\end{equation}%
with $\mathbf{K}_{r}=(\mathbf{k}_{r1}+\mathbf{k}_{r2})/2$ and $\Delta 
\mathbf{k}_{r}=\mathbf{k}_{r2}-\mathbf{k}_{r1}$, where $E'_0$ is the
amplitude of the electric field and where ${\bf k_{r1}}$ and  ${\bf k_{r2}}$ 
are related to ${\bf k_1}$ and ${\bf k_2}$ respectively by Eq. (1). 

The relatively long wavelength cosine function in Eq. (\ref{2-plane}) moves
in the NIM with a velocity 
\begin{equation}
\mathbf{v}_{gr}=(\Delta \omega /|\Delta \mathbf{k}_{r}|^{2})(\Delta k_{x}{\ 
\hat{\mathbf{x}}}+\Delta k_{rz}{\hat{\mathbf{z}}}),
\end{equation}%
assuming that $|{\bf\Delta k}|<<|{\bf K}|$.
From the above expression, it is evident that $v_{grx}>0$ if 
$\Delta k_{x}>0$. Since $\omega _{1}<\omega _{2}$, we have 
$0<n(\omega _{1})\leq n(\omega_{2})$ and 
$n_{r}(\omega _{1})<n_{r}(\omega _{2})<0$ by the requirement of
causality which requires $d(n\omega )/d\omega >0$. One has $%
k_{r1z}^{2}-k_{r2z}^{2}=k_{r1}^{2}-k_{r2}^{2}+k_{2x}^{2}-k_{1x}^{2}>0$.
Since $k_{rz}=-|k_{rz}|$, $v_{grz}>0$. The group refraction is indeed
positive \cite{Valanju}. This is due to the simple fact that $v_{grx}>0$ if 
$v_{gx}>0$. Proper dispersion will only give $v_{grz}>0$ since the energy
should propagate away from the interface. But we shall see that the above
picture is not true for the energy flow.

Let us determine the average Poynting vector $\langle \mathbf{S}_{r}\rangle$. 
Using the magnetic field corresponding to $\mathbf{E_{r}}$%
of Eq. (8), $\mathbf{H}_{r}=E_{0}\sum_{j=1}^{2}(k_{rjz}{\ \hat{\mathbf{x}}}%
-k_{jx}{\hat{\mathbf{z}}})e^{i\mathbf{k}_{rj}\cdot \mathbf{r}-i\omega
_{j}t}/\omega _{j}$, $<{\bf S_r}>$ is found to be given by 
\begin{equation}
\left\langle \mathbf{S}_{r}\right\rangle =-{\frac{1}{2}}(1+\cos \Delta \phi
_{r})|E'_{0}|^{2}\sum_{j=1}^{2}\Big({\hat{\mathbf{x}}}{\frac{k_{jx}}{\omega
_{j}}}+{\hat{\mathbf{z}}}{\frac{k_{rjz}}{\omega _{j}}}\Big)
\end{equation}%
where $\Delta \phi _{r}=\Delta \mathbf{k}_{r}\cdot \mathbf{r}-\Delta \omega t
$. Since $k_{rz}<0$, one has $S_{x}<0$ and $S_{z}>0$. Thus, contrary to the
refraction of the cosine function in Eq. (\ref{2-plane}), the Poynting
vector is directed in the negative refraction direction, i.e. refracts
negatively.

We shall now demonstrate that by including more plane waves in our group,
one can get negative refraction of the group. Actually, just one more 
plane wave can achieve
that. Thus, let us include three plane waves, whose wave vectors form
a triangle, rather than being parallel. Let the magnitudes of the wave
vectors be $k$, $k+\delta k_{1}$, $k+\delta k_{2}$, and their angles with
the normal to the interface, $\theta $, $\theta +\delta \theta _{1}$, $%
\theta +\delta \theta _{2}$. Inside the PIM or the NIM, we have 
\begin{eqnarray}
E &=&e^{i(k_{x}x+k_{z}z-\omega t)}\Big(1+\exp [i(u-ct)\delta k_{1}+ivk\delta
\theta _{1}]  \nonumber \\
&&+\exp [i(u-ct)\delta k_{2}+ivk\delta \theta _{2}]\Big),  \label{E-3wv}
\end{eqnarray}
with $u=x\sin \theta +z\cos \theta $ and $v=x\cos \theta -z\sin \theta $ for
the PIM and $u=x\sin \theta +az$ and $v=x\cos \theta +bz$ for the NIM. Then
the lines whose equations are $u$=constant and $v$=constant are 
perpendicular for the PIM. Here
use has been made of the following expansion $k_{rz}(k+\delta k,\theta
+\delta \theta )\approx k_{rz}+a\delta k+bk\delta \theta $ with $a=k(\sin
^{2}\theta +{c|n_{r}|/\upsilon _{r}})/|k_{rz}|,b=k\sin 2\theta /2|k_{rz}|.$
The condition for maximum intensity for the quantity in brackets, the 
long wavelength envelope of the packet, is
determined by the equations $(u-ct)\delta k_{1}+vk\delta \theta
_{1}=2m_{1}\pi ,(u-ct)\delta k_{2}+vk\delta \theta _{2}=2m_{2}\pi $, whose
solution in the PIM is $x=(c_{1}\sin \theta +c_{2}\cos \theta )+\sin \theta
\>ct,z=(c_{1}\cos \theta -c_{2}\sin \theta )+\cos \theta \>ct$ with $%
c_{1}=2\pi (m_{2}\delta \theta _{1}-m_{1}\delta \theta _{2})/(\delta \theta
_{1}\delta k_{2}-\delta \theta _{2}\delta k_{1}),c_{2}=2\pi (m_{1}\delta
k_{2}-m_{2}\delta k_{1})/(k(\delta \theta _{1}\delta k_{2}-\delta \theta
_{2}\delta k_{1}))$, which are clearly only defined for $\delta k_{1}/\delta
k_{2}\neq \delta \theta _{1}/\delta \theta _{2}$.

Inside the NIM, the solution for the location of the intensity maxima is $%
x=(c_{2}a-c_{1}b-bct)/(a\cos \theta -b\sin \theta ),z=(c_{1}\cos \theta
-c_{2}\sin \theta +\cos \theta \>ct)/(a\cos \theta -b\sin \theta )$. From
the expressions for $a$ and $b$ under Eq. (\ref{E-3wv}), one has $a,b>0$ and 
$a\cos \theta -b\sin \theta >0$. Then $x(t)$ and $z(t)$, $dx/dt<0$ and $%
dz/dt>0$. Thus the \emph{refraction will be negative}. Let the angles of the
line $u=$constant and $v=$constant in the NIM with the $z$-axis be $\alpha $
and $\beta$ respectively. Then one has $\tan \alpha =-a/\sin \theta ,\quad \tan
\beta =-b/\cos \theta =k_{x}/k_{rz}$. So one always has $\pi /2<\alpha
<\beta <\pi$ inside the NIM. From the above expressions, one can see that
the maxima move in the $\beta $ direction, that is, anti-parallel to $%
\mathbf{k}_{r}$. The group velocity in NIM is given by 
\begin{equation}
\mathbf{v}_{gr}=-\upsilon _{r}{\hat{\mathbf{k}}_{r}}.
\end{equation}
This velocity is independent of how the incident wave packet is constructed.
The refraction of a group constructed from 4 plane waves is shown in Fig.\ref%
{Fig3}. The arguments presented above demonstrate that for any \emph{group
consisting of 3 or more plane waves whose wavevectors are not collinear, the
group refraction is negative}.

\begin{figure}[htbp]
\center{
\includegraphics [angle=0,width=3.6in]{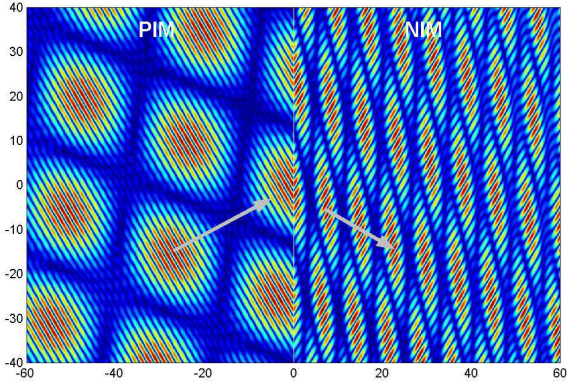}}
\vskip -.6cm
\caption{Electric field $\Re E$ of negative refraction of 4 plane waves with
wave vector magnitudes $k-\protect\delta k$,$k$,$k+\protect\delta k$,$k$,
and incident angles, $\protect\theta $, $\protect\theta -\protect\delta 
\protect\theta $, $\protect\theta $, $\protect\theta +\protect\delta \protect%
\theta $, respectively. The center wave number is $k=\protect\sqrt{5}$ with
incident angle $\protect\theta =\protect\pi /6$, $\protect\delta k=0.2$, and 
$\protect\delta \protect\theta =\protect\pi /45$. Up to the first order
approximation, the electric field, Poynting vector, and the moment of this
group of plane waves are $E_{r}=2e^{i\protect\phi _{r}}(\cos \protect\varpi %
+\cos \protect\delta \protect\phi _{r})$, $\left\langle \mathbf{S}%
_{r}\right\rangle =-2(\cos \protect\varpi +\cos \protect\delta \protect\phi %
_{r})^{2}\mathbf{k}_{r}/\protect\omega $, $\mathbf{P}_{r}^{\mathrm{cell}}=-2A%
\mathbf{k}_{r}/\protect\omega $, respectively.}
\label{Fig3}
\end{figure}

While the simulations in Fig.\ref{Fig3} clearly show that the intensity
maxima refract negatively, the normal to the planes in which these intensity
maxima lie are directed in a positive refraction direction. Thus, if one
were to imagine smoothing out all intensity variation in the planes, the
planes would appear to refract in a positive direction. We believe that this
is a remnant of the positive refraction of the planes of intensity maxima
(the cosine function in Eq. (\ref{2-plane})) found for the interference
patern for the two plane wave example of Ref. \cite{Valanju}. When there 
are only two plane waves, this is the only group motion that we see in the 
NIM since for a group consisting of two plane waves, there are no intensity 
variations in these planes. 

One can also look at the energy flow which is represented by the Poynting
vector. For three wave vectors with wave vector magnitudes $k-\delta k$, $k$%
, $k-\delta k$, and the angles with the normal, $\theta -\delta \theta $, $%
\theta $ , $\theta +\delta \theta $ respectively, the magnetic field can
also be calculated and hence the Poynting vector up to the first order in
both $\delta k$ and $\delta \theta $ is 
\begin{eqnarray}
\left\langle \mathbf{S}_{r}\right\rangle  &=&-{\frac{1}{2}}(1+4\cos
^{2}\varpi +4\cos \varpi \cos \delta \phi _{r})\mathbf{k}_{r}/\omega  
\nonumber \\
&&-\sin \varpi \sin \delta \phi _{r}\>(\cos \theta {\hat{\mathbf{x}}}+b{\hat{%
\mathbf{z}}})k\delta \theta /\omega   \nonumber \\
&&+2\cos ^{2}\varpi \>(a-k_{rz}/k)\delta k{\hat{\mathbf{z}}}/\omega 
\end{eqnarray}%
where $\varpi =(bz+\cos \theta \>x)k\delta \theta $ and $\delta \phi
_{r}=(az+\sin \theta \>x)\delta k-\delta \omega t$. Here, $\langle \mathbf{S}%
_{r}\rangle $ is not localized; rather it forms a lattice. A unit cell is
defined as the region in which $\varpi $ changes by $\pi $ and $\delta \phi
_{r}$ changes by $2\pi $, as is obvious from the expression for $E_{r}$ or $%
\langle \mathbf{S}_{r}\rangle $. The area for each unit cell in NIM is $%
A=2\pi ^{2}/(a\cos \theta -b\sin \theta )k\delta k\delta \theta $. Instead
of integrating over all space which will diverge, one can calculate the
electromagnetic momentum for each cell. Ignoring higher order terms in $%
\delta k$ and $\delta \theta $, we get 
\begin{equation}
\mathbf{P}_{r}^{\mathrm{cell}}=-3A(1+2\delta k/3k)\mathbf{K}_{r}/2\omega 
\end{equation}%
with $\mathbf{K}_{r}=\mathbf{k}_{r}-2(\sin \theta {\hat{\mathbf{x}}}+a{\hat{%
\mathbf{z}}})\delta k/3$, the average of the three wave vectors which make
up the group.

A packet constructed from a finite number of plane waves will always give a
collection of propagating wave pulses with the area of the unit cell
inversely proportional to $\delta k$ and $\delta\theta$. For the above
localized waves made of finite number of plane waves, the group velocity $%
\mathbf{v}_{gr}$ is parallel to $\mathbf{P}_r$ and anti-parallel to the
average wave vector $\mathbf{K}_r$.

In this paper, we have shown that for any localized wave packet, the
refraction at an interface between a PIM and an NIM is always negative. 
As pointed out earlier, 
it is essential for a correct treatment of this problem to use wave packets 
which are localized in all directions since the EM field from
any physical source is a localized wave packet. 

This work was supported by NSF-0098801, the Air Force Research Laboratories
and the Department of Energy.



\end{document}